\documentclass[twocolumn,preprintnumbers,amsmath,amssymb,superscriptaddress]{revtex4-2}

\usepackage{graphicx}
\usepackage{bm}
\usepackage{color}
\usepackage{ulem}
\usepackage{tikz}

\begin{document}

\title{Interaction of soliton gases in deep-water surface gravity waves}

\author{Loic Fache}
\affiliation{Univ. Lille, CNRS, UMR 8523 - PhLAM -
  Physique des Lasers Atomes et Mol\'ecules, F-59 000 Lille, France}
\author{F\'elicien Bonnefoy}
\affiliation{Nantes Universit\'e, \'Ecole Centrale Nantes, CNRS, LHEEA, UMR 6598, F-44 000 Nantes, France}
\author{Guillaume Ducrozet}
\affiliation{Nantes Universit\'e, \'Ecole Centrale Nantes, CNRS, LHEEA, UMR 6598, F-44 000 Nantes, France}
\author{Fran\c{c}ois Copie}
\affiliation{Univ. Lille, CNRS, UMR 8523 - PhLAM -
  Physique des Lasers Atomes et Mol\'ecules, F-59 000 Lille, France}
\author{Filip Novkoski}
\affiliation{Universit\'e Paris Cit\'e, CNRS,  MSC, UMR 7057, F-75 013 Paris, France}
\author{Guillaume Ricard}
\affiliation{Universit\'e Paris Cit\'e, CNRS,  MSC, UMR 7057, F-75 013 Paris, France}
\author{Giacomo Roberti}
\affiliation{Department of Mathematics, Physics and Electrical Engineering, Northumbria University, Newcastle upon Tyne, NE1 8ST, United Kingdom}
\author{Eric Falcon}
\affiliation{Universit\'e Paris Cit\'e, CNRS,  MSC, UMR 7057, F-75 013 Paris, France}
\author{Pierre Suret}
\affiliation{Univ. Lille, CNRS, UMR 8523 - PhLAM -
  Physique des Lasers Atomes et Mol\'ecules, F-59 000 Lille, France}
\author{Gennady El}
\affiliation{Department of Mathematics, Physics and Electrical Engineering, Northumbria University, Newcastle upon Tyne, NE1 8ST, United Kingdom}
\author{St\'ephane Randoux}
\email{stephane.randoux@univ-lille.fr}
\affiliation{Univ. Lille, CNRS, UMR 8523 - PhLAM -
  Physique des Lasers Atomes et Mol\'ecules, F-59 000 Lille, France}

\date{\today}

\begin{abstract}
  Soliton gases represent large random soliton ensembles in physical systems that display integrable dynamics at the leading order. We report hydrodynamic experiments in which we investigate the interaction between two ``beams'' or ``jets''  of soliton gases having nearly identical amplitudes but opposite velocities of the same magnitude. The space-time evolution of the two interacting soliton gas jets is recorded in a $140-$m long water tank where the dynamics is described at leading order by the focusing one-dimensional nonlinear Schr\"odinger equation. Varying the relative initial velocity of the two species of soliton gas, we change their interaction strength and we measure the macroscopic soliton gas density and velocity changes due to the interaction. Our experimental results are found to be in good quantitative agreement with predictions of the spectral kinetic theory of soliton gas despite the presence of perturbative higher-order effects that break the integrability of the wave dynamics.
\end{abstract}


\maketitle

\section{Introduction}\label{sec:intro}

Soliton gas (SG) is a concept in statistical mechanics and nonlinear physics that has been originally introduced by V. Zakharov in 1971 \cite{Zakharov:71} as a large  random ensemble of interacting solitons of the Korteweg-de Vries (KdV) equation. In the original Zakharov's model, the KdV SG is {\it diluted} with all solitons being individually discernible in the physical space where they occupy random positions and have random amplitudes. The emergent  dynamics of  SG on a macroscopic (hydrodynamic) scale, significantly larger than the characteristic soliton width, is determined by the fundamental properties of the ``elementary'' interaction between individual solitons. Owing to the integrable nature of the KdV equation  soliton collisions are pairwise (multi-particle effects are absent) and elastic, so that the interaction does not change the soliton amplitudes and velocities but produces only the additional position (phase) shifts \cite{Novikov_book}. 

In ref. \cite{Zakharov:71} Zakharov introduced the kinetic equation for a non-equilibrium {\it diluted} gas of weakly interacting solitons of the KdV equation. The Zakharov kinetic equation was generalized to the case of a dense (strongly interacting) KdV SG in ref. \cite{el_thermodynamic_2003}.  The kinetic theory of SG for the focusing one-dimensional nonlinear Schr\"odinger equation (1D-NLSE) has been developed in refs. \cite{GEl:05, GEl:19}.

Due to the presence of an infinite number of conserved quantities, random ensembles of nonlinear waves in integrable systems do not reach the thermodynamic equilibrium state characterized by an equipartition of energy leading to the so-called Rayleigh-Jeans distribution of the modes. Consequently, the properties of SGs are very different compared to the properties of classical gases whose particle interactions are non-elastic. The question of the thermodynamic properties  of SGs is addressed by  invoking {\it generalized hydrodynamics} (GHD), the hydrodynamic theory of many-body quantum and classical integrable systems \cite{Doyon:18, doyon_geometric_2018,Bonnemain:22,Bouchoule:22}.

It is well known that a comprehensive description of solitons and their interactions in physical systems described by integrable equations like the KdV equation or the 1D-NLSE is achieved within the framework of the celebrated inverse scattering transform (IST) method \cite{Ablowitz:73,Remoissenet_book,yang2010nonlinear,Zabusky:65,Novikov_book}. In the IST method, each soliton is parametrized by a discrete eigenvalue of a linear spectral problem associated with the nonlinear wave equation under consideration \cite{Prins:22}. The fundamental property of integrable dynamics is isospectrality, i.e. the preservation of the soliton spectrum (the eigenvalues) under evolution.

The central quantity of interest in SG theory is the density of states (DOS), which represents the statistical distribution over the spectral (IST) eigenvalues. The spectral kinetic description of non-uniform (non-equilibrium) SGs involves the continuity equation for the DOS (associated with the isospectrality condition) and the equation of state defining the effective velocity of a tracer soliton inside a SG, which differs from its  velocity in the ``vacuum'' due to the pairwise interactions  with other solitons in the gas, accompanied  by the position/phase shifts.

Despite the significant developments of the SG theory \cite{carbone_macroscopic_2016, shurgalina_nonlinear_2016, girotti_rigorous_2021, girotti_soliton_2023, kachulin_soliton_2020, GEl:11,GEl:19,Gelash:19,GEl:21, congy_soliton_2021, Bonnemain:22}, the experimental and observational results related to SGs are quite limited \cite{Costa:14,Hassaini:17,Cazaubiel:18,Marcucci:19,Redor:19,Redor:21,Suret:20}. In recent works, it has been shown that SGs with controlled and measurable DOS can be generated in laboratory experiments made with deep-water surface gravity waves \cite{Suret:20}. An important step towards the quantitative verification of the spectral kinetic theory of SG has recently been made in optical fiber experiments where the refraction of a soliton by a dense soliton gas has been demonstrated \cite{Suret:23}. In this experiment, the velocity change experienced by the tracer soliton in its interaction with an optical SG has been found to be in good quantitative agreement with the results of the spectral kinetic theory of SG. 

In this article, we report further experiments to investigate the physical validity of the kinetic theory of SG. Instead of considering the interaction between a single tracer soliton and a SG like in ref. \cite{Suret:23}, we examine the interaction between two SG ``beams'' or ``jets''   in hydrodynamic experiments performed with deep-water surface gravity waves. By the SG jet we mean a SG having a narrow distribution of the discrete IST eigenvalues around some given point in the complex spectral plane.  Sometimes such special SG's are called  monochromatic, with the DOS modeled by the Dirac delta-function. Mathematically, the introduction of a  DOS  in the form of a linear superposition of several delta-functions (the ``polychromatic'' ansatz) leads to a significant simplification of the kinetic equation and the availability of  analytical solutions describing various SG interactions \cite{GEl:05, GEl:11, carbone_macroscopic_2016, carr_2023}.

In our experiments, we consider  the interaction of two monochromatic SG jets that are configured to have equal amplitudes and opposite velocities. In physical space, each jet has the form of a large ensemble of individual solitons, with all the solitons having nearly the same amplitude and velocity. This configuration  has been considered theoretically in ref. \cite{GEl:05} by formulating an appropriate  Riemann problem for  the SG kinetic equation. In this specific setting  the DOS in the interaction region represents a linear superposition of two  delta-functions, which reduces the SG kinetic equation to two quasilinear partial differential equations of hydrodynamic type.  As shown in \cite{GEl:05, carbone_macroscopic_2016, GEl:21} the Riemann problem for the resulting two-component hydrodynamic system  admits a simple weak solution consisting of three constant states (for each component) separated by two propagating contact discontinuities.  This solution, in particular, describes  the component density and velocity changes resulting from the nonlinear interaction between two SG jets. In this paper, we present hydrodynamic experiments where the theoretical predictions from the spectral kinetic theory of SG are verified with good accuracy, further confirming its physical validity.

This article is organized as follows. In Sec. \ref{sec:background}, we present the theoretical background from kinetic theory of SGs, which is necessary to describe the interaction between SG jets in the framework  of the focusing 1D-NLSE. We illustrate the theoretical results with numerical simulations of the reduced kinetic equation describing the evolution in space and time of the densities of the two SG components. In Sec. \ref{sec:simu_nls}, we show how the IST method can be used to realize the implementation of two interacting SG jets in direct numerical simulations of the 1D-NLSE. In Sec. \ref{sec:exp}, we report our experimental results and compare them with the predictions of the kinetic theory. 

\section{Theoretical background}\label{sec:background}

In this section, we provide a brief summary of the theoretical results from the SG theory that are relevant to the description of the interaction between two spectrally ``monochromatic''  SG jets. More details about this special class of SGs can be found in refs. \cite{GEl:05, GEl:11, carbone_macroscopic_2016,  congy_soliton_2021, Kamchatnov:23}. We also illustrate the main theoretical results from the kinetic theory of SGs with some numerical simulations of the simplified SG kinetic equation describing the ``two-jet'' interactions.

\subsection{Analytical results from the spectral kinetic theory of SG}\label{sec:analytic}

We consider nonlinear wave systems described by the integrable focusing 1D-NLSE that reads 
\begin{equation}\label{eq:NLSE_adim}
  i \psi_t+\psi_{xx}+2|\psi|^2 \psi=0.
\end{equation}
The fundamental soliton solution of Eq. (\ref{eq:NLSE_adim}) parameterized by the the complex IST eigenvalue $\lambda=\alpha+i\gamma$ ($\alpha \in \mathbb{R}$, $\gamma \in \mathbb{R^+}$) reads
\begin{equation}\label{eq:NLSE_soliton}
  \psi(x,t)=2 \gamma \frac{ \exp [ -2i \alpha x -4i(\alpha^2-\gamma^2)t -i\phi_0  ]}{\cosh[2 \gamma(x+4 \alpha t-x_0)]},
\end{equation}
where $x_0$ and $\phi_0$ represent the initial position and phase parameters. 
The real part of the eigenvalue $\lambda$ encodes the velocity  $-4\alpha$ of the soliton in the $(x,t)$ plane, while the imaginary part  determines its amplitude $2 \gamma$ (as a matter of fact, the  IST spectrum of \eqref{eq:NLSE_soliton} also includes the complex conjugate  $\lambda^*=\alpha-i\gamma$). 

In the spectral kinetic theory of 1D-NLSE SG, the DOS represents the distribution $f(\lambda;x,t)$ over the spectral eigenvalues, so that $f d \alpha d \gamma dx$ is the number of soliton states found at time $t$ in the element of the 3D phase space $[\alpha, \alpha + d \alpha] \times [\gamma, \gamma + d \gamma]\times [x,  x+ dx]$. Due to the isospectrality condition associated with the integrable nature of Eq. (\ref{eq:NLSE_soliton}), the space-time evolution of the DOS $f(\lambda;x,t)$ is governed by the continuity equation
\begin{equation}\label{eq:continuity}
  \frac{\partial f}{\partial t} + \frac{\partial (sf)}{\partial x} = 0, 
\end{equation}
where $s=s(\lambda;x,t)$ represents the transport velocity of a tracer soliton inside a SG. For the focusing 1D-NLSE, the equation of state connecting the SG transport velocity with the DOS $f(\lambda;x,t)$ reads
\begin{equation}\label{eq:state}
  \begin{split}
  s(\lambda; x,t)=-4 \Re(\lambda) \, + \, \frac{1}{\Im(\lambda)} \iint\displaylimits_{\Lambda^+} \ln \left| \frac{\mu- \lambda^*}{\mu- \lambda} \right| \\
   [s(\lambda;x,t)-s(\mu;x,t)] \, f(\mu;x,t) d\xi d\zeta 
  \end{split}
\end{equation}
where $\mu=\xi + i \zeta$ and $\Lambda^+ \subset \mathbb{C^+} \setminus  i \mathbb{R^+}$ represents the 2D compact domain or 1D curve in the upper complex half-plane where the discrete eigenvalues parametrizing the SG of interest are located (it is sufficient to consider only  the upper half-plane  due to the c.c. (Schwarz) symmetry of the soliton spectrum).

Eqs. (\ref{eq:continuity}), (\ref{eq:state}) form the general kinetic equation for the focusing 1D-NLSE SG (see \cite{GEl:05, GEl:19}). It is a nonlinear integro-differential equation describing the evolution in space and time of the SG DOS $f(\lambda,x,t)$. The system (\ref{eq:continuity}), (\ref{eq:state}) can be considerably simplified if it is assumed that the SG is composed of a finite number of ``monochromatic'' components, or SG jets, each characterized by a DOS in the form of a Dirac delta-function.  
Here we concentrate on the two-component case involving two species of solitons with identical amplitudes and opposite velocities. The corresponding DOS has the form  
\begin{equation}\label{eq:dos}
  f(\lambda;x,t)=\rho_1(x,t) \, \, \delta(\lambda-\lambda_1) + \rho_2(x,t) \, \, \delta(\lambda-\lambda_2)
\end{equation}
with $\lambda_1=-\alpha+i\gamma$ and $\lambda_2=\alpha+i\gamma$. Here $\rho_{1,2}(x,t)$ are the SG component densities.  

Under the ansatz (\ref{eq:dos})  Eqs. (\ref{eq:continuity}), (\ref{eq:state}) reduce to the following ``two-jet'' hydrodynamic system  \cite{GEl:05}
\begin{equation}\label{eq:two_beam}
  \begin{split}
  \frac{\partial \rho_1(x,t)}{\partial t} + \frac{\partial (s_1(x,t) \, \rho_1(x,t))}{\partial x} = 0,\\
  \frac{\partial \rho_2(x,t)}{\partial t} + \frac{\partial (s_2(x,t) \, \rho_2(x,t))}{\partial x} = 0,
  \end{split}
\end{equation}
with the component transport velocities given by
\begin{equation}\label{eq:velocities}
\begin{split}
  s_1 = 4 \alpha \frac{1-\kappa (\rho_1 - \rho_2)}{1-\kappa (\rho_1 + \rho_2)},\\
  s_2 = -4 \alpha \frac{1+\kappa (\rho_1 - \rho_2)}{1-\kappa (\rho_1 + \rho_2)}.
 \end{split}
\end{equation}
Here $\kappa$ is the  interaction parameter  
\begin{equation}\label{eq:kappa}
  \kappa=\frac{1}{2 \gamma} \ln \left( 1 + \frac{\gamma^2}{\alpha^2} \right), 
\end{equation}
which represents the space shift due to the collision between two individual solitons with spectral parameters $\lambda_1$ and $\lambda_2$ \cite{Zakharov:72}.


As observed in \cite{GEl:05} (see also \cite{Kamchatnov:23}) system \eqref{eq:two_beam}, \eqref{eq:velocities} is equivalent to the  so-called Chaplygin gas equations,  the system of isentropic gas dynamics with  the equation of state $p= - {A}/{\rho}$, where $p$ is the pressure, $\rho$ is the gas density and $A>0$ is a constant. The Chaplygin gas equations   occur in certain theories of cosmology (see e.g. \cite{bento_generalized_2002}) and are also equivalent to the 1D Born-Infeld equation  arising in nonlinear electromagnetic field theory \cite{born_foundations_1934,whitham}.
The fundamental property of system \eqref{eq:two_beam},  \eqref{eq:velocities} is its linear degeneracy. Indeed, upon introducing the dependent variables $s_{1,2}(x,t)$ instead of $\rho_{1,2}(x,t)$ in \eqref{eq:two_beam} one arrives at the diagonal system
\begin{equation}
\frac{\partial s_1}{\partial t} + s_2 \frac{\partial s_1}{\partial x} = 0, \quad \frac{\partial s_2}{\partial t} + s_1 \frac{\partial s_2}{\partial x} = 0,
\end{equation}
with the characteristic velocities not depending on the corresponding Riemann invariants. Linear degeneracy of system  \eqref{eq:two_beam},  \eqref{eq:velocities} implies the principal absence of wavebreaking effects accompanied by the classical shock formation with the only admissible singularities being contact discontinuities \cite{rozhdestvenskii_systems_1983}.

\begin{figure}
  \includegraphics[width=0.45\textwidth]{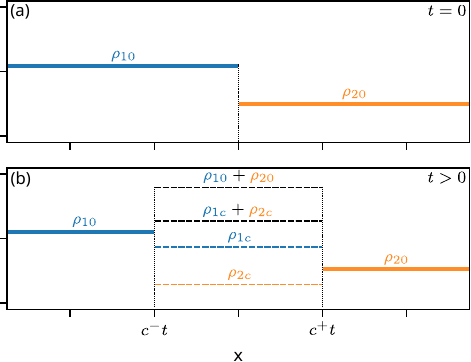}
  \caption{(a) Initial condition (Eq. \ref{eq:initial_densities}) for the Riemann problem for the two-jet hydrodynamic system (Eqs. \eqref{eq:two_beam}, \eqref{eq:velocities}) and (b) schematic of the solution given by Eq. \eqref{eq:sol}.}
  \label{fig1}
    \end{figure}

Following ref. \cite{GEl:05}, we use system  \eqref{eq:two_beam},  \eqref{eq:velocities} to describe the collision between two  SG jets with spatially uniform DOS's: $\rho_{10}\delta(\lambda - \lambda_1)$ and $\rho_{20} \delta(\lambda - \lambda_2)$---that are spatially separated at initial time. The corresponding initial condition for  Eq. (\ref{eq:two_beam}) has the form
\begin{equation}\label{eq:initial_densities}
  \begin{split}
  \rho_1(x,0)=\rho_{10}, \qquad \rho_2(x,0)=0 \qquad \text{for} \qquad x<0, \\
  \rho_1(x,0)=0, \qquad \rho_2(x,0)=\rho_{20} \qquad \text{for} \qquad x>0,  \\
  \end{split}
\end{equation}
and it is schematically shown in Fig.~\ref{fig1}(a).

This is a Riemann or ``shock-tube'' problem for the system of hydrodynamic conservation laws \eqref{eq:two_beam}.  Its solution, schematically shown in Fig.~\ref{fig1}(b), consists of three constant states for $(\rho_1, \rho_2)$ separated by two contact discontinuities \cite{GEl:05}:
\begin{equation}
\label{eq:sol}
(\rho_1(x,t), \rho_2(x,t)) =
\begin{cases}
(\rho_{10}, 0)  &x <c^- t,\\
(\rho_{1c}, \rho_{2c}) & c^-t \leq x  < c^+t\\
 (0, \rho_{20}) &c^+t \leq x,
\end{cases}\quad
\end{equation}
where the values of the component densities $\rho_{1c}, \rho_{2c}$ in the interaction region and the velocities  $c^{-}$ and $c^+$ of the contact discontinuities are found from the Rankine-Hugoniot conditions to be (see \cite{GEl:05} for details)
\begin{equation}\label{eq:densities_interaction}
\begin{split}
  \rho_{1c} =  \frac{\rho_{10}(1-\kappa \rho_{20})}{1-\kappa^2 \rho_{10} \rho_{20}},\\
  \rho_{2c} =  \frac{\rho_{20}(1-\kappa \rho_{10})}{1-\kappa^2 \rho_{10} \rho_{20}}.
\end{split}
\end{equation}

\begin{equation}\label{eq:velocities_interaction}
\begin{split}
c^- = s_{2c} = -4 \alpha \frac{1+\kappa (\rho_{1c} - \rho_{2c})}{1-\kappa (\rho_{1c} + \rho_{2c})}, \\
c^+=s_{1c} = 4 \alpha \frac{1-\kappa (\rho_{1c} - \rho_{2c})}{1-\kappa (\rho_{1c} + \rho_{2c})}.
 \end{split}
\end{equation}

\begin{figure}
  \includegraphics[width=0.45\textwidth]{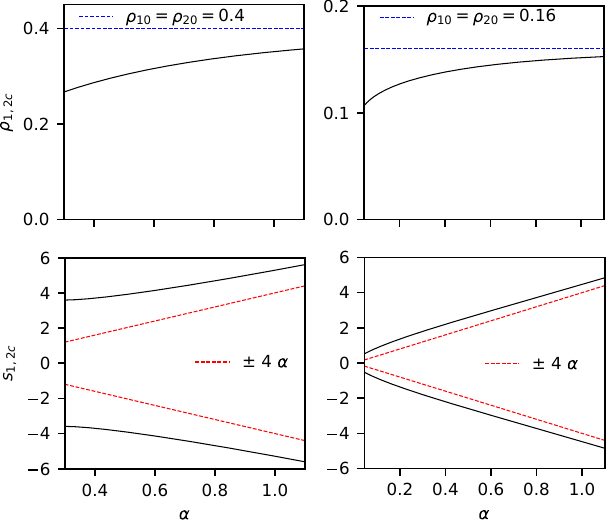}
  \caption{Evolution of the densities $\rho_{1,2c}$ and of the velocities $s_{1,2c}$ of the interacting SG jets as a function of $\alpha$, the parameter determining the relative velocity of the two jets. The plots in the left (resp. right) column are computed from Eq. (\ref{eq:densities_interaction}) and Eq. (\ref{eq:velocities_interaction}) with parameters that describe the densities of the non-interacting SGs being $\rho_{10}=\rho_{20}=0.4$ (resp. $\rho_{10}=\rho_{20}=0.16$) and $\gamma=1$. The red dashed lines represent the free velocities $\pm 4 \alpha$ of the non-interacting SGs ($\kappa=0$).} 
  \label{fig2}
  \end{figure}

 One should note that the denominators in \eqref{eq:densities_interaction}, \eqref{eq:velocities_interaction} never vanish due to fundamental restriction related to the notion of critical, or condensate, DOS (see \cite{GEl:19}). Moreover, it is not difficult to show that the interaction between the two SGs results in a ``dilution'' of each of the two species, i.e.  $\rho_{1c}< \rho_{10}$, $\rho_{2c} < \rho_{20}$.

Fig.~\ref{fig2} shows  the densities $\rho_{1,2c}$ and the velocities $s_{1,2c}$ in the interaction region as functions of $\alpha$, which is the parameter that determines the relative velocities of the two SG species. The parameter $\gamma$ determining the amplitude of the solitons has been fixed to unity. The initial densities are taken to be $\rho_{10}=\rho_{20}=0.4$ in the left column and $\rho_{10}=\rho_{20}=0.16$ in the right column. For the values of $\alpha$ that are large enough ($\alpha \gtrsim 0.7$), the interaction parameter $\kappa$ is relatively small ($\kappa \lesssim 1$) and the kinetic theory predicts that the density changes in the interaction region are relatively small ($\rho_{1,2c} \sim \rho_{1,20}$). On the other hand, the interaction between the two species increases when their initial relative velocity is small ($\alpha \lesssim 0.5$). This results in the density changes that are more significant for smaller values of $\alpha$.

The dashed red lines with slopes $\pm 4 \alpha$ in Fig.~\ref{fig2} represent the velocities that each species of SG would have in the $(x,t)$ plane without any interaction with the other one ($\kappa=0$ in Eq. (\ref{eq:velocities_interaction})). The black lines in the bottom row of Fig.~\ref{fig2} indicate the velocities $s_{1,2c}$ that are taken by each species as the result of the interaction with the other one. The comparison between the right and left columns shows that the velocity changes are more important when the initial density of the SGs is large. One of the goals of this paper is to compare the theoretical curves presented in Fig.~\ref{fig2} with results of physical experiments, see Sec. \ref{sec:exp}. 

We now present two series of numerical simulations where we verify the weak solution \eqref{eq:sol} by (i) numerically solving the two-jet kinetic equation  \eqref{eq:two_beam}  and (ii) performing direct simulations of the 1D-NLSE \eqref{eq:NLSE_adim}. Both simulations are performed for the initial data relevant to the physical experiments to be discussed in Section~\ref{sec:exp}.

\subsection{Numerical simulations of the kinetic equations}\label{sec:numeric_kinetic}

\begin{figure*}[!t]
  \includegraphics[width=1.02\textwidth]{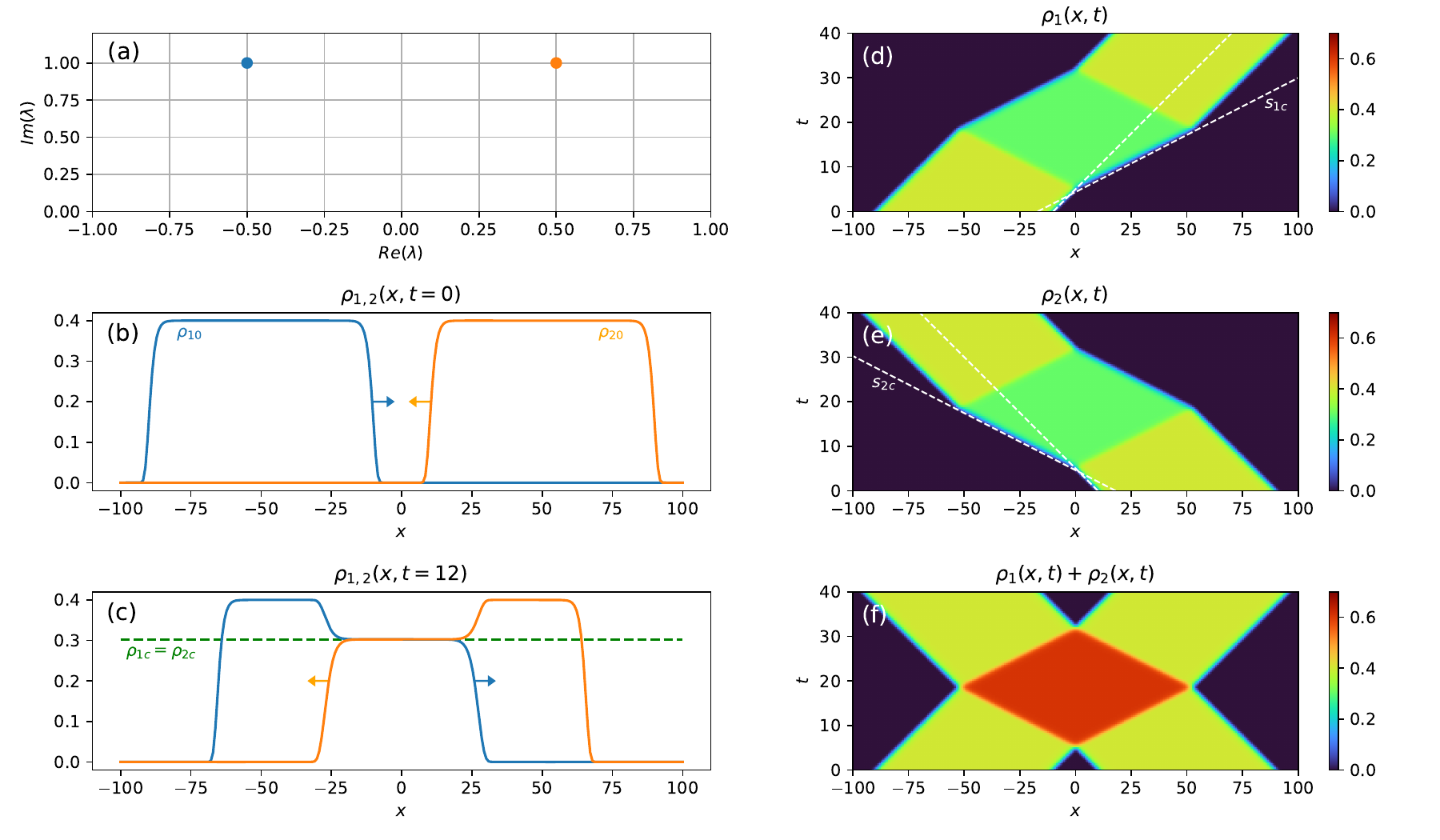}
  \caption{Numerical simulations of the ``two-beam'' kinetic equations (Eq. \ref{eq:two_beam}) showing the interaction between two jets of SGs. (a) Spectral (IST) parameters of the two interacting SGs, with the DOS being defined by Eq. (\ref{eq:dos}) with $\lambda_1=-0.5+i$ and $\lambda_2=0.5+i$. (b) Initial distribution of the densities $\rho_{1,2}(x,t=0)$. (c) Numerically computed distribution of the densities at $t=12$. The green dashed line represents the densities in the interaction region that are computed using Eq. (\ref{eq:densities_interaction}) with $\rho_{10}=\rho_{20}=0.4$ ($\alpha=0.5$, $\gamma=1$). (d) Space-time evolution of the density $\rho_1(x,t)$. The region in green is the interaction region where the density has decreased from $\rho_{10} = 0.4$ to $\rho_{1c} \sim 0.302$. (e) Same as (d) but for the second species $\rho_2(x,t)$. (f) Space-time evolution of the sum of the densities $\rho_1(x,t)+\rho_2(x,t)$ showing that the total density has increased in the interaction region despite the individual densities have decreased. }
  \label{fig3}
\end{figure*}

Fig.~\ref{fig3} shows numerical simulations of the kinetic equation illustrating the theoretical results presented in Sec. \ref{sec:analytic}. We consider two SG jets with the DOS being given by Eq. (\ref{eq:dos}) and $\lambda_{1,2}= \mp 0.5+i$ ($\alpha=0.5$, $\gamma=1$), as shown in Fig.~\ref{fig3}(a). We have numerically integrated the ``two-jet'' kinetic equations (\ref{eq:two_beam}) using a standard pseudo-spectral method where the space derivatives are computed in Fourier space. To avoid numerical problems associated with the finite size of the numerical box and the discontinuities of the initial condition used in the analytical calculations of ref. \cite{GEl:05} (Eq. (\ref{eq:initial_densities})), the initial condition taken in our numerical simulations is composed of two boxes of large extents and uniform initial densities $\rho_{10}=\rho_{20}=0.4$, as shown in Fig.~\ref{fig3}(b).

Fig.~\ref{fig3}(d)(e) show the space-time evolutions of the densities $\rho_{1,2}(x,t)$ of the two SG jets that are initially separated and start interacting from $t\sim 5$. As a result of the interaction the density of each species falls from $\rho_{10} = \rho_{20} =0.4$ to $\sim 0.302$, see the color scale that changes from yellow to green in Fig.~\ref{fig3}(d)(e). The numerical value of the densities computed in the interaction region is in perfect agreement with theoretical predictions, as shown in Fig.~\ref{fig3}(c) where the green dashed line represents the densities $\rho_{1c}=\rho_{2c}$ that are computed using the analytical expressions given by Eq. (\ref{eq:densities_interaction}).

In addition to the density changes due to the interaction, Fig.~\ref{fig3}(d)(e) show that the velocity changes found in numerical simulations are also in perfect agreement with theoretical predictions, see white dashed lines parallel to the boundaries of the SGs and associated with velocities $s_{1c}\sim 3.898$ and $s_{2c}\sim -3.898$ that are given by Eq.~(\ref{eq:velocities_interaction}). Finally, Fig.~\ref{fig3}(f) shows that despite the density of each species decreases due to the interaction, the total density $\rho_{1c}+\rho_{2c}$ of the SG in the interaction region  is larger than the individual densities $\rho_{1,2}(x,t)=\rho_{10,20}$ of each gas outside the interaction region. At the same time, $[\rho_{1c}+\rho_{2c}] < [\rho_{10}+\rho_{20}]$, i.e. the SG component interaction leads to the overall dilution compared to the non-interacting  two-component gas. This feature has already been pointed out in ref. \cite{GEl:05}.

Summarizing, the kinetic theory of SG predicts that the interaction between two monochromatic SG jets having opposite mean velocities but identical mean amplitudes results in density and velocity changes that are illustrated in Figs.~\ref{fig2},~\ref{fig3}. Our goal in this paper is to perform a hydrodynamic experiment to quantitatively verify these theoretical predictions. Before moving to experimental results, we present in Sec. \ref{sec:simu_nls} direct numerical simulations of the 1D-NLSE corresponding  to the numerical simulations of the two-jet kinetic equations shown in Fig.~\ref{fig3}. 

\begin{figure*}[!t]
  \includegraphics[width=1\textwidth]{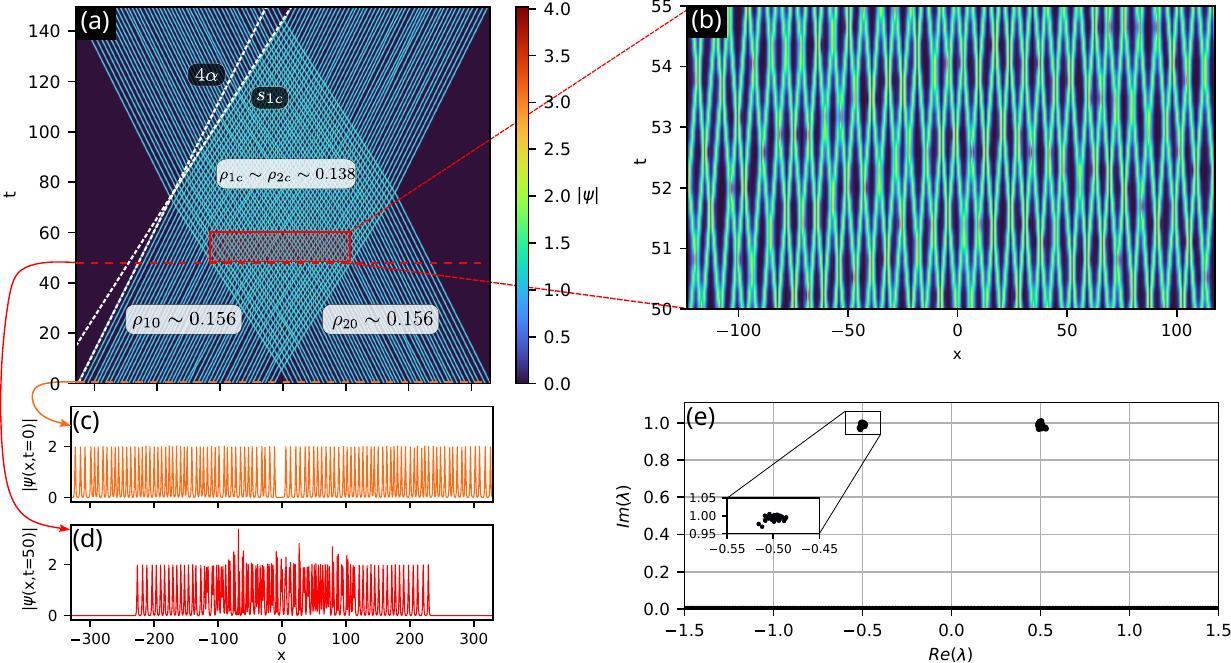}
  \caption{Numerical simulations of Eq. (\ref{eq:NLSE_adim}) with the initial condition being under the form of two ``monochromatic'' beams of SGs with opposite velocities. At initial time, each beam of SG is composed of $50$ solitons with nearly identical amplitudes and opposite velocities ($\alpha=0.5$, $\gamma=1$). (a) Space time plot showing velocity and density changes arising from the interaction between the two SGs. (b) Enlarged view of the interaction region showing microscopic dynamics and multiple elastic collisions between individual solitons. (c) Modulus $|\psi(x,t=0)|$ of the initial condition. (d) Modulus of the field at time $t=48$. (e) Discrete IST spectrum of the field composed of two separate clouds of $50$ eigenvalues centered around $\lambda_{1,2}=\mp 0.5 + i$.}
    \label{fig4}
\end{figure*}

\section{Interacting soliton gas jets in numerical simulations of the 1D-NLSE}\label{sec:simu_nls}

In this Section, we show how the IST method can be used to realize the implementation of two interacting jets of SGs not in numerical simulations of the kinetic equations but in numerical simulations of the 1D-NLSE .

A nonlinear wave field $\psi(x,t)$ satisfying Eq. (\ref{eq:NLSE_adim}) can be characterized by the so-called scattering data (the IST spectrum). For localized wave fields decaying to zero as $x \rightarrow \infty$ the IST spectrum consists of a discrete part related to the soliton content and a continuous part related to the dispersive radiation. A special class of solutions, the N-soliton solutions (N-SSs), exhibits only a discrete spectrum consisting of N complex-valued eigenvalues $\lambda_n$, $n = 1, ..., N$ and $N$ complex parameters $C_n =|C_n| e^{j \phi_n}$, called norming constants, defined for each $\lambda_n$. The complex discrete eigenvalues encode the amplitudes and velocities of the solitons while the norming constants encode their phases and ``positions'' in physical space \cite{Novikov_book}. 

Using a recursive algorithm based on the Darboux transform \cite{Garcia:19}, we have built a N-SS of Eq. (\ref{eq:NLSE_adim}) with $N=100$. The discrete eigenvalues associated with this N-SS are partitioned into two random sets, each being linked to a given SG. The first (resp. second) SG is parameterized by $50$ eigenvalues that are randomly distributed in an uniform way within a small square region of the complex plane centered around $\lambda_1=-0.5+i$ (resp. $\lambda_2=0.5+i$), see Fig.~\ref{fig4}(e). Following the approach described in ref. \cite{Gelash:18,Roberti:21}, we have synthesized the SG by implementing the Darboux recursive scheme in high precision arithmetics, a requirement due to the large number of solitons composing the wave field. The wave field has been synthesized at $t=0$ and a standard NLSE solver based on a pseudo-spectral method has been used to compute the space-time evolution at longer time, as shown in Fig.~\ref{fig4}(a). At initial time, the two SGs are separated without any spatial overlap between the two species, see Fig.~\ref{fig4}(c). Each of the two SGs is composed of $50$ solitons having approximately the same amplitude while being individually discernible. The random nature of each gas can be appreciated in physical space through the fact that the distance between neighboring solitons is not fixed but random.

Let us emphasize that the two SGs that we have realized are as dense as possible. The Darboux method is a recursive transformation scheme where a “seeding solution” of the focusing 1D-NLSE is used as a building block for the construction of a higher-order solution through the addition of one discrete eigenvalue. The Darboux transform machinery produces N-SS in such as way that the smaller the distance between the eigenvalues, the greater the physical separation between the solitons in physical space. For our SG, the mean distance in physical space between neighboring solitons of each species is therefore determined by the size of the square regions where the discrete eigenvalues are located, see Fig.~\ref{fig4}(e). However the mean distance between solitons not only depends on the distance between the eigenvalues $\lambda_n$ but also on the norming constants $C_n$. In Fig.~\ref{fig4}, the SG has been made as dense as possible by setting the moduli $|C_n|$ of the norming constants to unity and by distributing uniformly their phases $\phi_n$ between $0$ and $2 \pi$, similarly to what has been done in ref. \cite{Suret:20}. The SG of Fig.~\ref{fig4} cannot be denser than it is but it could be diluted by randomly distributing the moduli of the norming constants over some interval having a nonzero extent.  

At time $t=0$ each of the two species constitutes a uniform SG whose density $\rho_0$ represents the number $n$ of solitons relative to the interval of length $l$ they occupy: $\rho_0=n/l$. In Fig.~\ref{fig4}(a)(c), the initial densities $\rho_{10}$ and $\rho_{20}$ of each of the two non-interacting species are $n/l \sim 50/320 \sim 0.156$, which is the highest possible for the spectral parameters that have been chosen (see Fig.~\ref{fig4}(e)). This means that the numerical results presented in this Section and their associated experimental results presented in Sec. \ref{sec:exp} must to be compared with theoretical predictions of the kinetic theory that are plotted in the right column of Fig.~\ref{fig2} where $\rho_{10}=\rho_{20}=0.16$.

Fig.~\ref{fig4}(a) shows that the interaction between the two species results in a ``dilution'' associated with a drop in the densities. In the center of the interaction region, at time $t \sim 75$, each of the two species containing $n=50$ solitons now occupy a spatial domain having an extent that has increased from $l \sim 320$ to $l' \sim 362$. This results in a decrease of the densities that fall from $\rho_{10}=\rho_{20} \sim 0.156$ to $\rho_{1c}=\rho_{2c}=n/l'=50/362 \sim 0.138$, in good quantitative agreement with the expressions (\ref{eq:densities_interaction}) obtained with the framework of the kinetic theory of SG. In addition to density changes, Fig.~\ref{fig4}(a) also shows that the interaction between the two species of SG leads to changes in their relative velocities. Simulations of the 1D-NLSE plotted in Fig.~\ref{fig4}(a) show that the mean velocity of the first species increases from $4 \alpha \sim 2$  to $s_{1c} \sim 2.57$ due to the interaction, once again in good quantitative agreement with the results from the kinetic theory (Eq. \ref{eq:velocities_interaction}).

Recent optical fiber experiments reported in ref. \cite{Suret:23} have investigated the interaction between an individual tracer soliton and a dense SG. It has been shown that the tracer soliton experiences an effective velocity change due to its interaction with the optical SG. The experimental features observed in this optical fiber experiment are qualitatively similar to the classical refraction phenomenon observed in ray optics at the interface between two dielectric media having different refractive indexes. Here, the space-time evolution shown in Fig.~\ref{fig4}(a) for two  SG jets is also reminiscent from ray optics with one beam/jet of SG being shifted in space but not due to the propagation in a medium with another refractive index, but due to the nonlinear interaction with another beam/jet of SG. Note that the velocity and density changes measurable for each species of SG at the macroscopic scale are the emergent effects due to the numerous elementary elastic collisions  between individual solitons occurring at the microscopic, soliton,  scale, as shown in Fig.~\ref{fig4}(b). 

\section{Experiments}\label{sec:exp}

\subsection{Experimental setup and generation of the initial wave field}\label{sec:exp_setup}

\begin{figure}
  \includegraphics[width=0.5\textwidth]{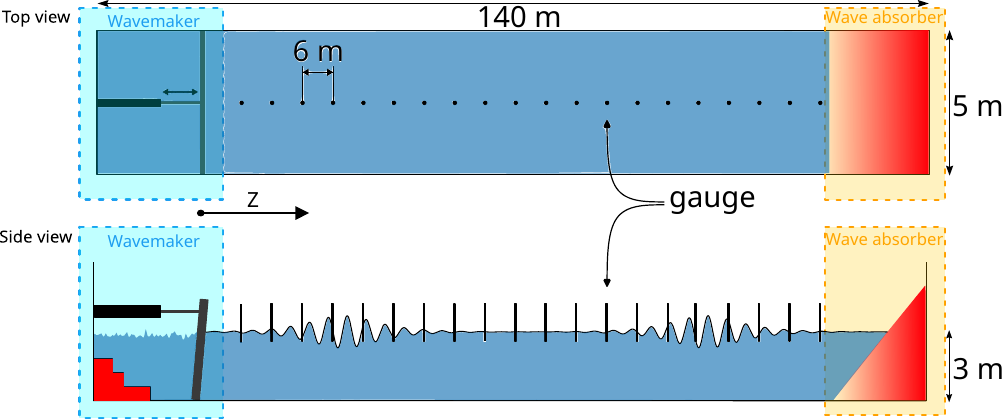}
  \caption{Schematic representation of the 1D water tank used in the experiments. $20$ wave elevation gauges are placed every $6$ meters, covering a measurement range of $114$ meters.}
  \label{fig5}
\end{figure}

The experiments have been performed in a wave flume at the Hydrodynamics, Energetics and Atmospheric Environment
Lab (LHEEA) in Ecole Centrale de Nantes (France). The flume which is $140$ m long, $5$ m wide and $3$ m deep is equipped with an absorbing beach that is approximately $8$ m long, see Fig.~\ref{fig5}. With the addition of pool lanes arranged in a W pattern in front of the beach the measured amplitude reflection coefficient is as low as $1\%$. Unidirectional waves are generated with a computer assisted flap-type wavemaker. As in the experiments reported in ref. \cite{Bonnefoy:20,Suret:20}, the setup comprises $20$ equally spaced resistive wave gauges that are installed along the basin at distances $Z_j=j \times 6$ m, $j=1,2,...20$ from the wavemaker located at $Z=0$ m. This provides an effective measuring range of $114$ m.

\begin{figure*}[!t]
  \includegraphics[width=1\textwidth]{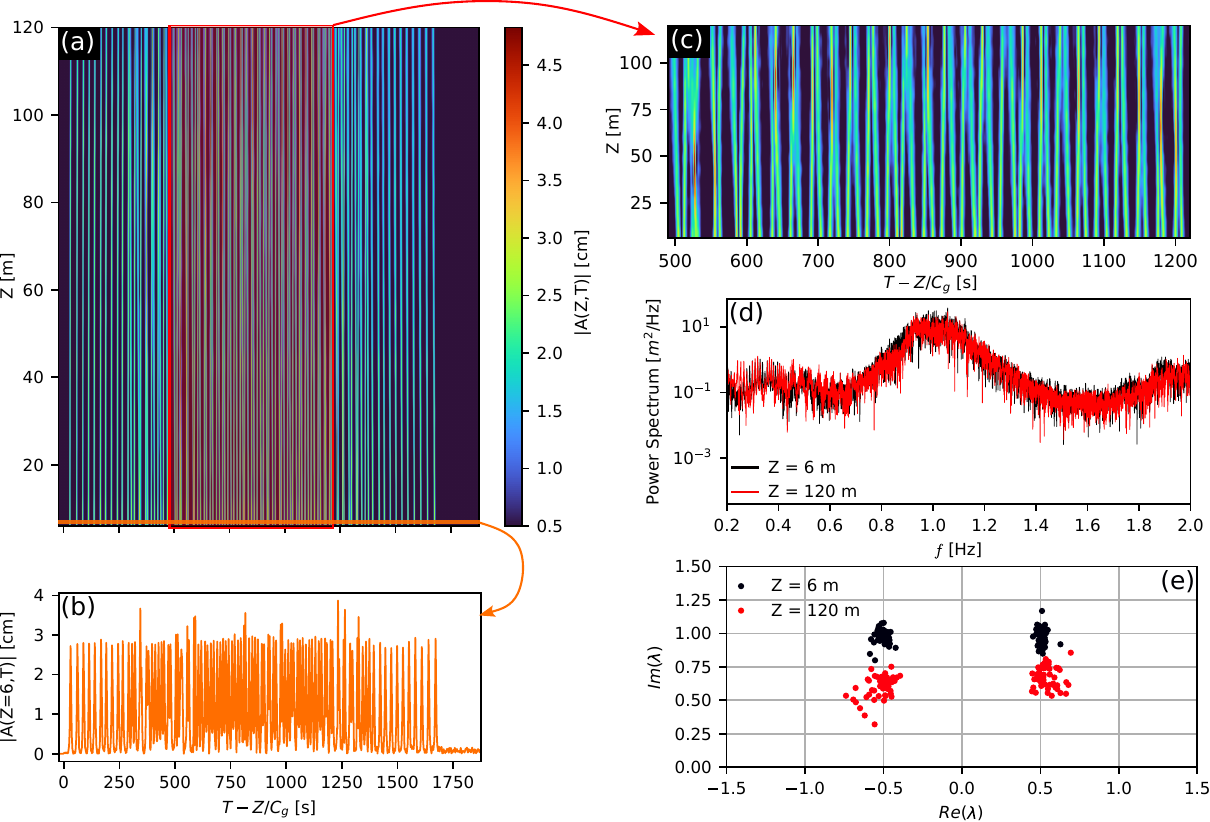}
  \caption{Experiments performed in the 140-m long water tank with two interacting SG jets, each being composed of $50$ solitons with spectral (IST) parameters $\alpha= \mp 0.5$ and $\gamma=1$. (a) Space-time evolution of the two ``monochromatic'' SG jets with the central region being the interaction region. In the two lateral regions of the space-time diagram, the two species of SGs propagate with opposite velocities without interacting. (b) Modulus of the envelope of the wave field measured by the first gauge at $Z=6$ m. (c) Enlarged view of the interaction region showing individual collision between solitons occurring at random positions inside the water tank. (d) Fourier power spectra of the elevation of the wave field measured at $Z=6$ m and at $Z=120$ m. (e) Discrete IST spectra of the envelope of the wave field measured at $Z=6$ m and at $Z=120$ m. Experiments are made for a carrier frequency $f_0=1.01$ Hz and a steepness $k_0 a \simeq 0.115$ ($L_{NL}\simeq20.3$ m). 
  }
  \label{fig6}  
\end{figure*}

In our experiments, the water elevation at the wavemaker reads $\eta(Z=0,T)=Re \left[ A_0(T)e^{i\omega_0 T} \right] $, where $\omega_0 = 2 \pi f_0$ is the angular frequency of the carrier wave. In all the experiments presented in our paper, the frequency of the carrier wave is set to $f_0=1.01$ Hz. $A_0(T)$ represents the complex envelope of the initial condition.  Our experiments are performed in the deep-water regime, and they are designed in such a way that the observed dynamics is described at leading order by the focusing 1D-NLSE
\begin{equation}\label{nlse_phys}
 \frac{\partial A}{\partial Z} + \frac{1}{C_g} \frac{\partial A}{\partial T}=  i \frac{k_0}{\omega_0^2}
\frac{\partial^2 A}{\partial T^2} + i  \beta k_0^3 |A|^2 A,  
\end{equation}
where $A(Z,T)$ represents the complex envelope of the water wave that changes in space $Z$ and in time $T$ \cite{osborne2010nonlinear}. $k_0$ represents the wavenumber of the propagating wave ($\eta(Z,T)=Re \left[ A(Z,T)e^{i( \omega_0 T - k_0 Z)} \right]$), which is linked to $\omega_0$ according to the deep water dispersion relation $\omega_0^2=k_0 g$, where $g$ is the gravity acceleration. $C_g=g/(2\omega_0)$ represents the group velocity of the wavepackets and $\beta \simeq 0.91$ is a dimensionless term describing the small finite-depth correction to the cubic nonlinearity \cite{Bonnefoy:20}. 

The first important step of the experiment consists in generating an initial condition $A_0(T)$ in the form of two ``monochromatic'' beams of SGs, as illustrated in Fig.~\ref{fig4}(c). To achieve this, we have to convert the dimensionless fields synthesized as initial conditions (see Sec. \ref{sec:simu_nls}) into physical units. Connections between physical variables of Eq. (\ref{nlse_phys}) and  dimensionless variables in Eq. (\ref{eq:NLSE_adim}) are given by $t=Z/(2L_{NL})$, $x=(T-Z/C_g)\sqrt{g/(2 L_{NL})}$ with the nonlinear length being defined as $L_{NL}=1/(\beta k_0^3 \, a^2)$, where $a$ represents the mean peak amplitude of solitons outside the interaction region ($a\simeq 2.8$ cm in all our experiments).

Numerical simulations of Fig.~\ref{fig4}(a) show that $\sim 140$ units of normalized time are needed for two beams of SGs to overlap, interact and separate. This large normalized evolution time corresponds to an unrealistic physical propagation distance over $280$ nonlinear lengths, with the nonlinear length $L_{NL}$ being typically around $\sim 20$ m in the experiments that we are dealing with \cite{Bonnefoy:20,Suret:20}. To take account for the fact that our hydrodynamical experiments cannot go beyond propagation distances longer than $\sim 6$ $L_{NL}$, we have designed our initial wavefield in such a way that it is composed of a total number of $100$ solitons with one central interaction region and two lateral regions where each species does not interact with the other, see Fig.~\ref{fig6}(b). Note that the SGs outside the interaction region are uniform with constant densities being equal to $\rho_{1,20}=0.156$.

\subsection{Space-time evolution, measurement of the Fourier and of the discrete IST spectra}\label{sec:exp_space_time}

Taking two beams of solitons with spectral (IST) parameters identical to those used to compute Fig.~\ref{fig4}, Fig.~\ref{fig6}(a) shows the space-time diagram reconstructed from the signals recorded by the 20 gauges. Note that our experiments deal with envelope solitons. The signal recorded by the gauges is therefore composed of a carrier wave at a frequency $f_0 \sim 1.01$ Hz that is slowly modulated by a solitonic envelope. The first step in processing the experimental data consists in removing the carrier wave and in computing the complex envelope $A(Z,T)$ of the measured wavefield, which is achieved by using standard Hilbert transform techniques \cite{osborne2010nonlinear}. The space-time diagram of Fig. ~\ref{fig6} is plotted in a reference frame moving at the mean group velocity $C_g$ of the two monochromatic SG jets. In this reference frame, the two SG jets have opposite velocities of the same magnitude. 

Fig.~\ref{fig6}(a) and ~\ref{fig6}(b) shows that the wavefield is composed of one central interacting region and two lateral regions where each species does not interact with the other. Fig. ~\ref{fig6}(c) is an enlarged view into the interaction region. It shows that, despite the relatively short propagation distance ($\sim 6 L_{NL}$) reached in the experiment, individual interactions occur between pairs of solitons at {\it random} propagation distances in the water tank. These paired interactions occurring at {\it microscopic} level are responsible for {\it macroscopic} density and velocity changes that are measurable and that will be discussed in Sec. \ref{sec:exp_densities}. 

Fig.~\ref{fig6}(d) shows the Fourier power spectra of the elevation of the wave fields that are measured at $Z=6$ m, close to the wavemaker and at $Z=120$ m, far from the wavemaker. The propagation of the generated SGs is not accompanied by any significant broadening of the Fourier power spectrum.

Fig.~\ref{fig6}(e) shows the discrete IST spectra measured at $Z=6$ m and at $Z=120$ m. The discrete IST spectrum measured at $Z=6$ m consists of two narrow clouds of eigenvalues centered around $\lambda_{1,2}=\mp 0.5 +i$, in accordance with the initial condition we have engineered, see Sec. \ref{sec:simu_nls}. Each cloud represents an ensemble of $50$ discrete eigenvalues, with each of these discrete eigenvalue being associated with one of the solitons that propagates in the water tank (see Fig.~\ref{fig6}(a)).

The discrete IST spectrum measured at $Z=120$ m (red points in Fig.~\ref{fig6}(e)) is not identical to the discrete IST spectrum measured at $Z=6$ m. This means that the experiment is not perfectly described by the {\it integrable} 1D-NLSE (Eq. (\ref{nlse_phys})) and that the space-time dynamics is weakly perturbed by higher order effects, a feature that we have already observed and discussed in some of our previous experiments \cite{Bonnefoy:20,Suret:20,Tikan:22}. A discussion about the higher-order effects breaking the integrability of the wave dynamics is given in the Appendix. The important point here is that the IST analysis reveals that two separate clouds, each containing $50$ eigenvalues, retain a finite and limited extension in the complex plane during the nonlinear evolution. As a result, we can now examine the extent to which the predictions of the kinetic theory of SG remain robust in an experiment that cannot be inherently described by an integrable equation. 

\subsection{Measurement of the densities and velocities of the hydrodynamic SGs}\label{sec:exp_densities}

\begin{figure}[!t]
  \includegraphics[width=0.5\textwidth]{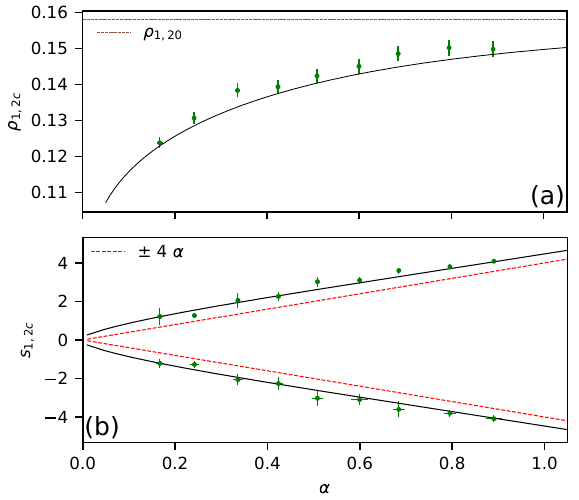}
  \caption{Comparison between the experiments and kinetic theory of SG. (a) Evolution of the densities $\rho_{1,2c}$ as a function of $\alpha$ in the interaction region. Green points represent experimental measurement points while the solid black line is computed using Eq. (\ref{eq:densities_interaction}) with $\rho_{1,20}=0.156$ and $\gamma=1$. (b) Same as (a) but for the velocities $s_{1,2c}$ of the interacting SGs. The red dashed lines represent the free velocities $\pm 4 \alpha$ of the non-interacting SGs. All the experiments have been made with $f_0=1.01$ Hz and for a steepness $k_0 a \simeq 0.115$. Error bars in (a) are associated with the uncertainty in the measurement of the space interval occupied by the SGs. Error bars in (b) represent the standard deviations associated with the velocity measurements, see Fig.~\ref{fig8}(b). }
    \label{fig7}
\end{figure}

In order to verify the predictions of the kinetic theory of SG, we carried out experiments to examine the validity of the velocity and density evolutions plotted in Fig.~\ref{fig2} using Eq. (\ref{eq:densities_interaction}) and (\ref{eq:velocities_interaction}). We have made an ensemble of $9$ experiments similar to the one depicted in Fig.~\ref{fig6} by changing the value of $\alpha$ between $\sim 0.2$ and $\sim 0.9$. In each of the $9$ experiments, we have used the IST-based methodology described in Sec. \ref{sec:simu_nls} to synthesize two interacting SGs with the parameter $\alpha$ being changed between $\sim 0.2$ and $\sim 0.9$, the parameter $\gamma$ being kept to one. We have recorded the associated space-time evolutions and we have checked that discrete IST spectra measured close and far from the wavemaker consist of two separate clouds composed of $50$ eigenvalues similar to those shown in Fig.~\ref{fig6}(e). 

The easiest macroscopic observables to measure in the experiment are the densities of each species $\rho_{1,2c}$ in the interaction region. To measure $\rho_{1,2c}$, we first convert the signals recorded in physical variables into dimensionless form by using relations given in Sec. \ref{sec:exp_setup}. Taking the dimensionless wavefield measured at the last gauge, we just count the number of solitons $n$ that we find for each species in the interaction region and we measure the space interval $l$ that these solitons occupy. As discussed in Sec. \ref{sec:simu_nls}, the measured density of the SGs is given by $\rho_{1,2c}=n/l$.

Fig.~\ref{fig7}(a) shows that we obtain a very good quantitative agreement between experiments and the kinetic theory of SG. The density of each species in the interaction region decreases from $\sim 0.15$ to $\sim 0.125$ when the value of $\alpha$ is changed from $\sim 0.9$ to $\sim 0.2$. In the experiment, there was no meaning in trying to further increase the interaction between the two SGs by reducing the value of $\alpha$ below $\sim 0.2$. For values of $\alpha$ smaller than $0.2$ the relative velocity of the two species is indeed so small that there is no significant interaction/collision between the two species over the relatively short propagation distance ($\sim 6 L_{NL}$) that is accessible in the experiment. 

Looking at the evolution pattern measured in the experiment (see Fig.~\ref{fig6}(a) and Fig.~\ref{fig6}(c)), it can considered at first sight that it is difficult, if not impossible, to determine the velocity of the SGs inside and even outside the interaction region. Following the approach proposed in ref. \cite{Redor:20} to separate right- and left-propagating solitons in a shallow water bidirectional SG, we have found that the Radon transform can be used to measure the velocities of the solitons in the space-time diagrams recorded in our experiments.

The two-dimensional Radon transform is an integral transform that maps a function to its integral over radial lines parameterized by an angle $\theta$ and by a distance $r$ to an origin point. The Radon transform $R(r,\theta)$ of the normalized space-time plots $|\psi(x,t)|$ recorded in the experiment reads:
\begin{equation}\label{eq:Radon}
  R(r,\theta)=\int \int |\psi(x,t)| \, \delta (x \cos \theta+t \sin \theta - r) dx dt
\end{equation}
where $\delta$ is the Dirac function. $r=\sqrt{x^2+t^2}$ is the distance to an origin point located in the center of $(x,t)$ domain. 

\begin{figure}[!t]
  \includegraphics[width=0.5\textwidth]{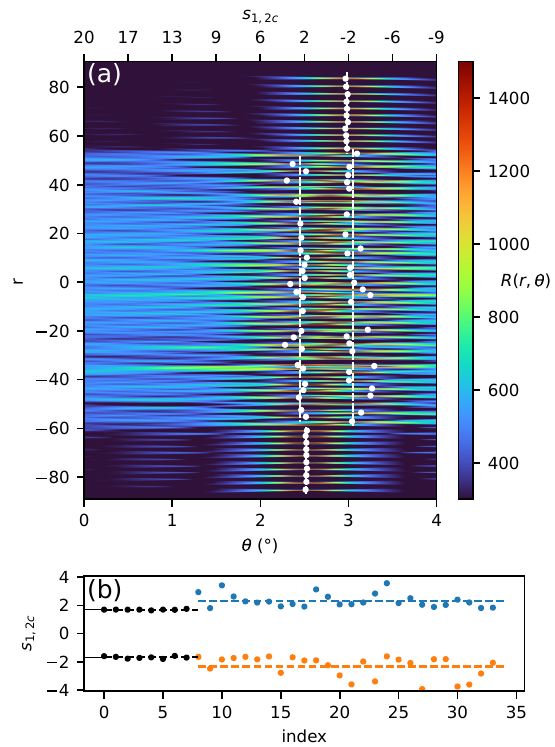}
  \caption{(a) Radon transform $R(r,\theta)$ of the experimental space-time diagram of Fig.~\ref{fig6}(a) for $\alpha= \mp 0.5$. The white points indicate the positions at which a maximum of the function $R(r,\theta)$ is found. (b) Simplified diagramatic view of the results obtained in (a) using the Radon transform. Two sets each containing $8$ free (non-interacting) solitons are found with mean velocities of $\sim 1.69$  and $\sim -1.69$ (black points). Two other sets each containing $25$ solitons are found in the interaction region with mean velocities of $\sim 2.34$ (blue points) and $\sim -2.34$ (red points).}
    \label{fig8}
\end{figure}

Fig.~\ref{fig8}(a) represents the Radon transform of the experimental space-time diagram of Fig.~\ref{fig6}(a), which has been normalized to dimensionless variables of Eq. (\ref{eq:NLSE_adim}) using variable transformations given in Sec. \ref{sec:exp_setup} ($\psi(x,t)=A/(a/2)$, $t=Z/(2L_{NL})$, $x=(T-Z/C_g)\sqrt{g/(2 L_{NL})}$). The Radon transform $R(r,\theta)$ immediately reveals the existence of several distinct classes of solitons being parameterized by their position $r$ relatively to the origin point and by an angle parameter $\theta$ related to their velocity in the $(x,t)$ plane. After applying a calibration procedure converting the angle parameter into a velocity parameter and after isolating the local maxima associated with each soliton in the Radon transform, we end up with the simple plot presented in Fig.~\ref{fig8}(b).

Fig.~\ref{fig8}(b) represents the velocities of the solitons that have been unambiguously detected using the Radon transform of the space-time diagram of Fig.~\ref{fig6}. Depending on the initial phase, position and precise velocity of each soliton, certain interaction patterns measured in physical space can produce signatures in the Radon transform, such as double peaks, which do not allow us to conclude unambiguously about the velocity taken by the solitons. These ambiguous measurement points are ignored and we finally obtain two sets, each containing not $50$ but $35$ solitons, for which we have a correct velocity measurement performed using the Radon transform. 

Fig.~\ref{fig8}(b) shows that $8$ isolated (non-interacting) solitons are detected with a velocity of $\sim 1.69$ and $8$ other non-interacting solitons are detected with a nearly opposite velocity of $\sim -1.69$. In the interaction region, the solitons with positive velocities have their mean velocity that increases to $\sim 2.34$ while the solitons with negative velocities have their mean velocity that decreases to $\sim -2.34$. Note that the dispersion of the velocities around the mean value is significantly larger in the interaction region as compared with the region where solitons do not interact. This is due to the fact that each paired interaction occurs at different random positions in the water tank, which results in a collection of microscopic interaction patterns associated with a larger dispersion of the values of velocities measured using the Radon transform. 

\begin{figure*}[!ht]
  \includegraphics[width=1\textwidth]{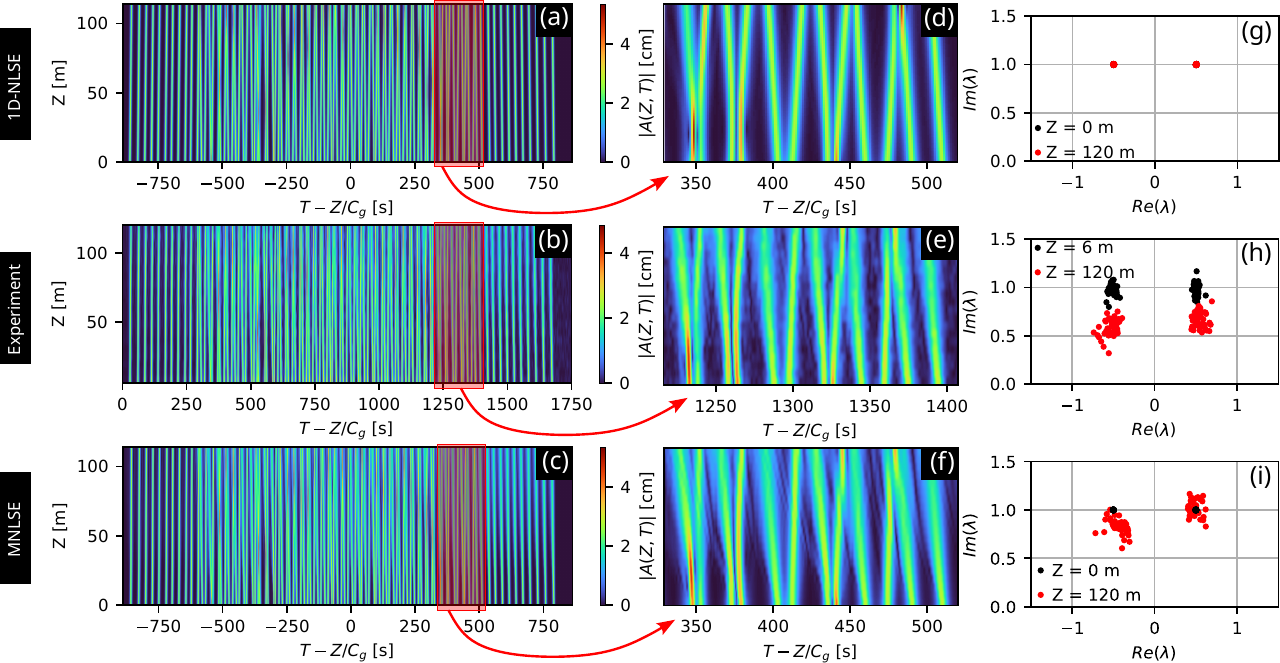}
  \caption{Comparison between experiments, numerical simulations of the focusing 1D-NLSE and of Eq. (\ref{dysthe_eq}) for the interaction between two jets of SG, each containing $50$ solitons. (a) Space-time diagram showing the space-time evolution described by the integrable 1D-NLSE. (c) Zoomed view into the interaction region. (g) Discrete IST spectra computed at $Z=0$ m and at $Z=120$ m. (b), (e) (h) Same as (a), (c), (g) but in the experiment. (c), (f), (i) Same as (a), (c), (g) but in numerical simulations of Eq. (\ref{dysthe_eq}). Parameters used in numerical simulations are $f_0=1.01$ Hz, $k_0 a=0.115$, $g=9.81$ m s$^{-2}$, $\beta=0.91$. }
    \label{fig9}
\end{figure*}

Fig.~\ref{fig7}(b) synthesizes all the measurements of the mean velocities that have been made in the interaction region on our $9$ experiments where the value of $\alpha$ has been changed between $\sim 0.2$ and $\sim 0.9$. Despite the existence of higher-order effects and the fact that the experiments are not perfectly described by the integrable 1D-NLSE, Fig.~\ref{fig7}(b) shows that the theoretical predictions of the kinetic theory in terms of velocity changes of the SGs are quantitatively well verified in the experiment.

\section{Conclusion}\label{sec:conclusion}

In this paper, we have reported hydrodynamic experiments in which we have investigated the interaction between two  SG jets having identical mean amplitude but opposite mean velocities. The two jets of interacting SGs are synthesized using the IST method. Their IST spectrum is composed of two clusters of discrete eigenvalues centered around two specific points of the complex spectral plane. We have recorded the space-time evolution of the interacting SGs in a $140-$m long water tank. We have varied the interaction strength between the two interacting species by changing their relative initial velocity. We have measured the macroscopic density and velocity changes due to the interaction between the two SG jets. Our experimental results are found to be in good quantitative agreement with predictions of the kinetic theory of SG despite the fact that the experiment is not perfectly described by the integrable 1D-NLSE.

We believe that our experimental results provide an important step towards the physical validation of the fundamental theoretical principles behind the spectral theory of SGs. We hope that they will stimulate new research in the field of statistical mechanics of nonlinear waves and integrable turbulence.

\appendix*
\section{Influence of higher-order effects}\label{sec:appendix}

In this Appendix, we use numerical simulations of the focusing 1D-NLSE and of a modified (non-integrable) 1D-NLSE to show the role of higher order effects on the observed space-time dynamics and on the discrete IST spectra of the two jets of interacting SGs. 

Following the work reported in ref. \cite{Goullet:11}, higher-order effects in 1D water wave experiments can be described by a modified NLSE written under the form of a spatial evolution equation
\begin{equation}\label{dysthe_eq}
  \begin{split}
  \frac{\partial A}{\partial Z} + \frac{1}{C_g} \frac{\partial A}{\partial T}= i \frac{k_0}{\omega_0^2}
  \frac{\partial^2 A}{\partial T^2} + i \beta k_0^3 |A|^2 A\\
  - \frac{k_0^3}{\omega_0} \left( 6|A|^2 \frac{\partial A}{\partial T}
  +2A\frac{\partial |A|^2}{\partial T}
  -2 i A \mathcal{H}\left[\frac{\partial |A|^2}{\partial T } \right] \right)
  ,
  \end{split}
\end{equation}
where $A(Z,T)$ represents the complex envelope of the wave field and $\mathcal{H}$ is the Hilbert transform defined by $\mathcal{H}[f]=(1/\pi) \int_{-\infty}^{+\infty} f(\xi)/(\xi-T)d\xi$.

When the last three terms are neglected in Eq. (\ref{dysthe_eq}), the integrable 1D-NLSE (\ref{nlse_phys}) is recovered. Figures \ref{fig9}(a)(d) show space-time diagrams in which the dynamics of interaction between the two jets of SG is governed by the integrable focusing 1D-NLSE. Fig.~\ref{fig9}(g) shows that the discrete IST spectra of the two interacting SGs consist of two narrow clouds centered around $\lambda_{1,2}=\mp 0.5+ i$. Because of the isospectrality condition underlying the integrable nature of the focusing NLSE, these IST spectra do no change with the propagation distance. 

Figures \ref{fig9}(c)(f) show space-time diagrams computed from the numerical integration of Eq. (\ref{dysthe_eq}) that takes into account the influence of higher-order terms. The space-time evolution plotted in Fig. \ref{fig9}(c)(f) is very similar to that observed in the experiments, see Fig. \ref{fig9}(b)(e). In particular, it can be clearly seen in Fig. \ref{fig9}(c) and in Fig. \ref{fig9}(f) that solitary waves emit some radiation, which in not the case in Fig. \ref{fig9}(b). The discrete IST spectra computed at $Z=6$ m and at $Z=120$ m show that the isospectrality condition is not fulfilled in the experiment and in the numerical simulation of Eq. (\ref{dysthe_eq}), compare Fig.~\ref{fig9}(h)(i) with Fig.~\ref{fig9}(g). Higher-order effects produce some spreading (or diffusion) of the discrete eigenvalues, which nevertheless remain confined to two distinct clouds.

\begin{acknowledgments}
This work has been partially supported  by the Agence Nationale de la Recherche through the StormWave (ANR-21-CE30-0009) and SOGOOD (ANR-21-CE30-0061) projects, the LABEX CEMPI project (ANR-11-LABX-0007), the Simmons Foundation MPS No. 651463 project, the Ministry of Higher Education and Research, Hauts de France council and European Regional Development Fund (ERDF) through the Contrat de Projets Etat-R\'egion (CPER Photonics for Society P4S). The authors would like to thank the Isaac Newton Institute for Mathematical Sciences for support and hospitality during the programme ``Dispersive hydrodynamics: mathematics, simulation and experiments, with applications in nonlinear waves''  when part of the work on this paper was undertaken. G. El's and G. Robertis's work was also supported by EPSRC  Grant Number EP/W032759/1.  G. Roberti thanks Simons Foundation for partial support.
\end{acknowledgments}






%

\end{document}